\documentclass[12pt,amsmath]{iopart}

\usepackage{cite}
\usepackage{graphicx}
\usepackage{mathrsfs}
\newcommand{\be}{\begin{eqnarray}}
\newcommand{\ee}{\end{eqnarray}}
\newcommand{\ra}{\rangle}
\newcommand{\la}{\langle}

\newcommand{\out}{{\rm out}}

\newcommand{\gp}{g^\prime}

\newcommand{\inv}{{\rm in}}
\begin{document}

\title[Capacity of black hole channel]{Classical information transmission capacity of quantum black holes}

\author{Christoph Adami}

\address{Department of Physics and Astronomy, Michigan State University, East Lansing, MI 48824}
\ead{adami@msu.edu}
\author{Greg Ver$\!$ Steeg}
\address{Information Science Institute, Viterbi School of Engineering, University for Southern California, Los Angeles, CA 90089}
\begin{abstract} The fate of classical information incident on a quantum
black hole has been the subject of an ongoing controversy in theoretical physics, 
because a calculation within the framework of semi-classical curved-space quantum field theory appears to show that the incident information is irretrievably lost, in contradiction  to time-honored principles such as time-reversibility and unitarity. 
Here, we show within this framework embedded in quantum communication theory that signaling from past to future null infinity in the presence of a Schwarzschild black hole can occur with arbitrary accuracy, and
thus that classical information is not lost in black hole dynamics. The calculation relies on a treatment
that is manifestly unitary from the outset, where probability
conservation is guaranteed because black holes stimulate the
emission of radiation in response to infalling matter. This stimulated 
radiation is {\em non-thermal} and contains all of the
information about the infalling matter, while Hawking radiation
contains none of it.
 \end{abstract}
\pacs{04.70Dy,03.67-a,03.65.Ud}
\maketitle

\section{Introduction}
Black hole evaporation poses a serious challenge to theoretical
physics because it is a problem at the intersection of general
relativity and quantum mechanics, precisely where our
understanding is the weakest. From the moment Hawking discovered
the eponymous radiation effect~\cite{Hawking1975}, many
researchers have concluded that black holes introduce an intrinsic
{\em irreversibility} into the laws of physics, together with an
intrinsic {\em unpredictability}~\cite{Hawking1976b}. The reasons
for such a radical departure from conventional physics at first
seem incontrovertible. Because the only identifiers (quantum
numbers) of a black hole that can be measured are its mass,
charge, and angular momentum, information that distinguishes an individual particle but is not encoded in these
quantum numbers would be lost after the particle disappeared
behind the horizon, making the dynamics strictly irreversible~\cite{Witten2012}.
This implies that, from the point of view of communication theory,
signaling between a sender that encodes information into a beam of
particles aimed at the horizon, for example, and a receiver at
rest outside the horizon at future null infinity, would be
fundamentally impossible.

If information is trapped within a black hole, we are faced with
another paradox, known as the {\em coherence
problem}~\cite{Wald1994}. Hawking radiation, while reducing the
mass of the black hole, results in a highly entangled quantum
state whose support is the joint Hilbert space spanned by the
horizon and future null infinity. While this state can be pure, the
quantum state at future null infinity (averaged over the horizon) is naturally mixed. If
black hole evaporation ultimately results in the disappearance of
the horizon, what happens to the quantum state at this point? Does
it remain mixed, or does it magically return to a pure state on
future null infinity? Standard arguments suggest that the latter cannot
happen if information remains trapped behind the horizon, because
black holes just before complete evaporation would be too small to encode
all of it. And even if they could, this much entropy could not be
radiated away instantly (see, e.g., the discussion in~\cite{Preskill1992}). 

Technically speaking, the coherence problem is different from the information paradox
because the former addresses the quantum state {\em after} black hole
evaporation, while the information paradox concerns the
possibility of signaling in the presence of a horizon.  
The two problems are intimately related, and often conflated so that a reference to the ``information paradox" may indicate either of these two problems. (They should perhaps more accurately be discussed as the ``classical information paradox" and the ``quantum information paradox", respectively.)  Here, we directly address the classical information paradox from the point of view of quantum communication theory. We show that the information transmission capacity for late-time particles sent into the black hole is non-zero, and that therefore classical signal states can be decoded with arbitrary accuracy by an observer suspended at future null infinity outside the black hole horizon. 

We derive the black hole channel capacity (for classical information transmission) within standard curved-space quantum field theory, that
is, in the semiclassical approximation where the back-reaction of the quantum dynamics on the metric field is ignored. Signaling is
made possible by the {\em stimulated emission} of radiation that must accompany absorption. The well known {\em spontaneous
emission} (Hawking radiation) turns out to provide the noise of the black hole channel, noise from which no information can be
extracted. In the following, we first calculate the joint quantum state of black hole and radiation when early-time
particles accrete during the formation of the black hole, and use standard quantum channel theory (see,
e.g.,~\cite{Preskill1998,Wilde2013}) to calculate the capacity of that channel. This capacity will turn out to be exactly equivalent to the capacity of the recently considered  ``Unruh" channel~\cite{Bradler2011} that considers communication between inertial signalers and accelerated observers. We then obtain the final state of late-time accreting particles (those sent into a black hole for the purpose of signaling), and calculate the capacity of this channel. Using a formalism developed by Sorkin~\cite{Sorkin1987}, we can write down an $S$-matrix that allows for the interaction of particles with the black hole with arbitrary absorption probability, and reproduces the known grey-body factors. The capacity we derive for black holes with arbitrary reflectivity will turn out to be positive for all black holes with positive mass. In the limit of perfectly absorbing black holes (no reflection) the late-time capacity turns out to be 
formally identical to the early-time capacity,  except for a renormalization of the parameter that describes the level of noise in the channel.

\section{Formalism}
In units where $\hbar=c=G=k=1$, a Schwarzschild black
hole in Einstein gravity has an entropy given by $S_{\rm BH}=4\pi M^2$, where $M$ is
the mass of the black hole, and a temperature $T=(8\pi
M)^{-1}$, the Hawking temperature. Black holes are formed in
stellar collapse of stars of sufficient mass, and can subsequently
accrete particles (see Fig.~1a). Through the process of virtual
pair formation near the event horizon (e.g., pair $P$ in Fig.~1a), with one of the pair's
particles disappearing behind the event horizon and the other
going off to future null infinity (${\mathscr I^+}$ in
Fig.~1), a black hole loses mass by providing it to the virtual
pair, which, in turn, goes on mass shell. The problem of loss of
predictability is easily seen in Fig.~1a, where different
trajectories of particles accreting onto the black hole are
outlined, and arbitrarily labelled. If the identity of these
labels is lost behind the horizon, we are faced with a many-to-one
mapping and concurrent loss of predictability (even if this
information is released at a later time) because the future state
of the universe cannot be ascertained given the labels of the
accreting particles. If the labels are destroyed in the
singularity, then we would be forced to assume that, to make
matters worse, the laws of physics allow non-unitary, and thus by
extension probability non-conserving, dynamics.

\begin{figure}[htbp]
\begin{center}
\includegraphics[width= 6.25cm,angle=0]{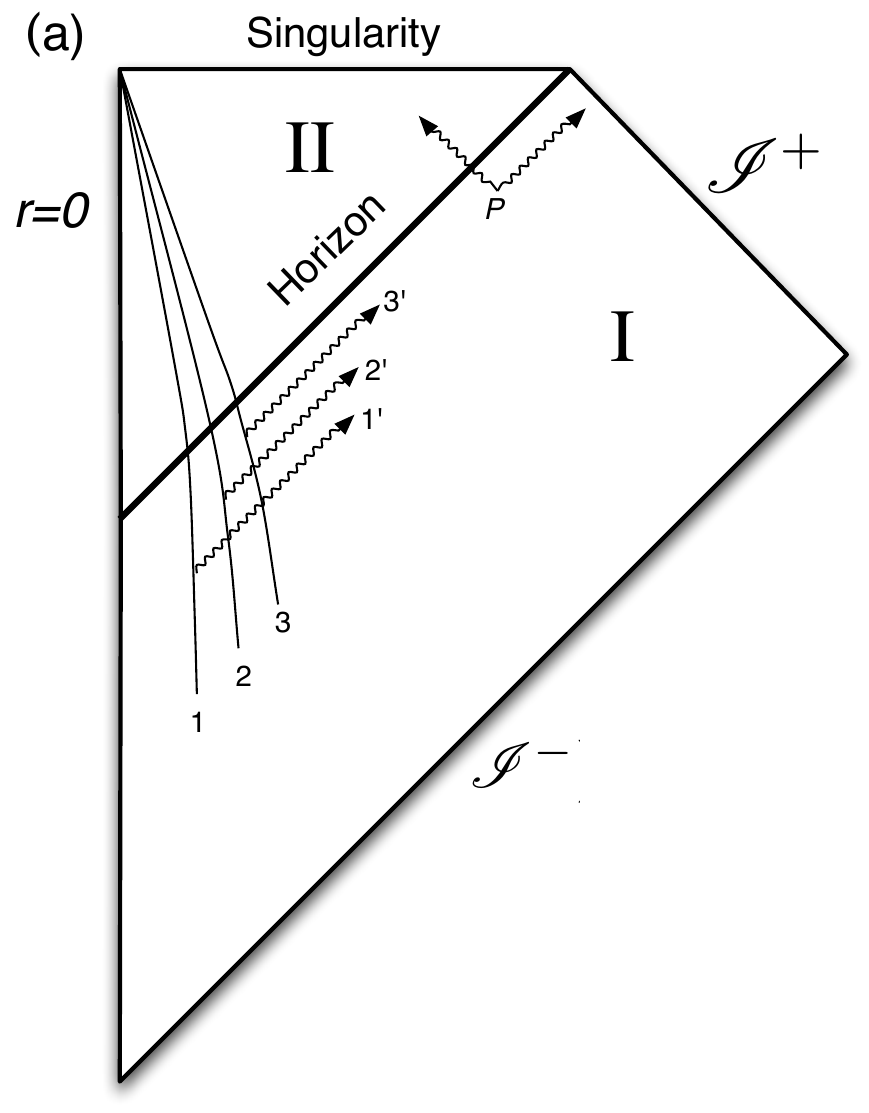}
\includegraphics[width= 6.25cm,angle=0]{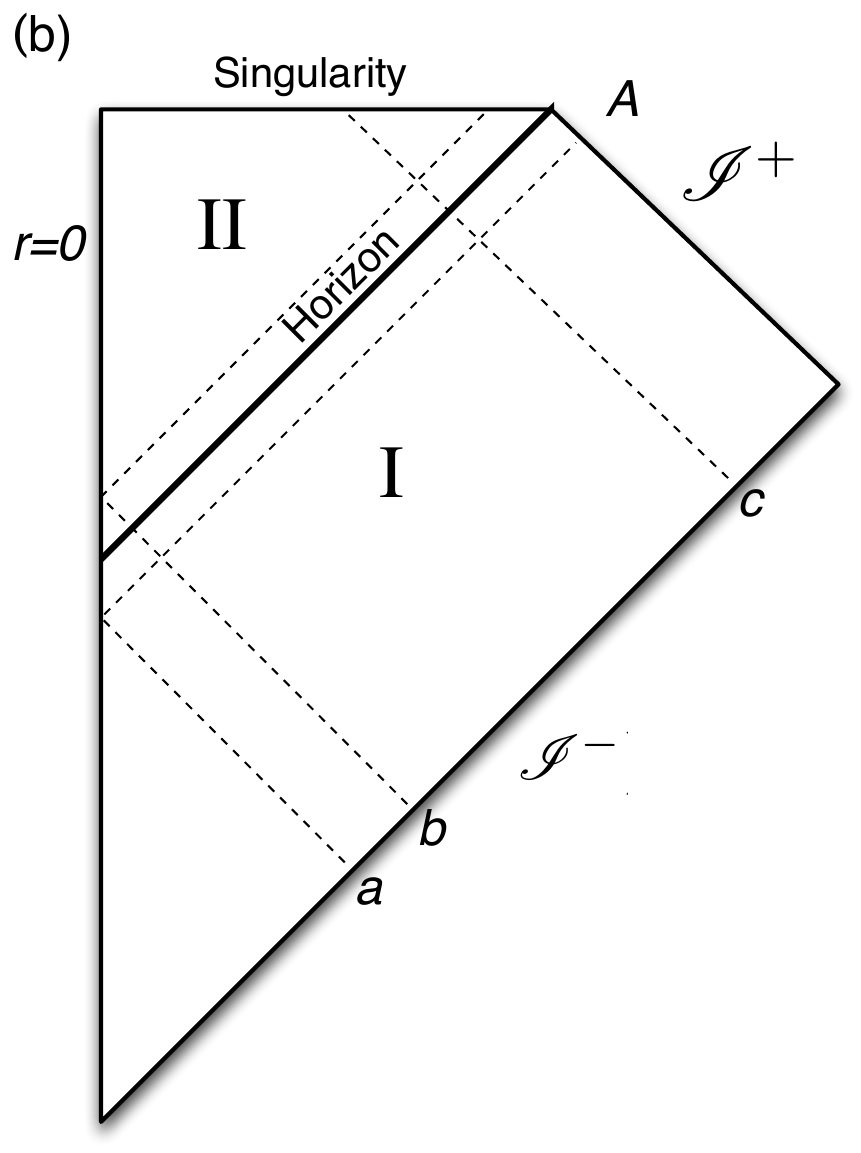}
\caption{(a) Penrose diagram of the spacetime of a black hole,
with accretion of arbitrarily labelled particles 1,2,3, from past null
infinity (${\mathscr I}^-$). Once the particles cross the event
horizon into region II, they are indistinguishable (leading to a
loss of information) unless they leave a signature outside (in
region I) via stimulated emission (labelled
$1^\prime,2^\prime,3^\prime$). (b) Modes $a$, $b$,
$c$ and $A$ are concentrated in a region of null infinity
indicated by the letter (note that $a$ and $b$ actually overlap on
${\mathscr I}^-$).  } \label{default}
\end{center}
\end{figure}

To introduce our notation, we begin by recapitulating the standard quantum theory of complex scalar fields in a gravitational background due to Hawking~\cite{Hawking1975}. The presence of a black hole's event horizon introduces a complication to quantum field theory not encountered in flat-space: different vacuum states can be defined that are inequivalent in terms of their particle content.  The so-called {\em Unruh} vacuum is non-trivial due to the separation of Schwarzschild space-time into two regions,
generally depicted as the ``outside" and ``inside" region, or region I and region II, respectively (Fig.~1). In the standard treatment, the operators that annihilate the curved-space vacuum, $A_k$ and $B_k$ ($A_k$
annihilates the ``outside" vacuum while $B_k$ annihilates``inside"), are related to the flat space operators $a_k$ and
$b_k$ by a Bogoliubov transformation~\cite{BirrellDavies1982}, so
 \begin{eqnarray}
A_k &=& e^{-iH}a_k e^{iH} = \alpha_k a_k - \beta_k b^\dagger_{-k}\label{bog1}\;,\\
B_k &=& e^{-iH}b_k e^{iH} = \alpha_k b_k - \beta_k a^\dagger_{-k}\;. \label{bog2} 
\end{eqnarray}
where $a_k^\dagger$ creates a particle of mode $k$ while
$a_{-k}$ annihilates an antiparticle of the same mode. The transformation provides a mapping from the trivial {\em Boulware} ``in"-vacuum $|0\ra_{\rm in}=|0\ra_a|0\ra_b$ (the Boulware vacuum equals the standard Minkowski vacuum far from the black hole) to the Unruh ``out"-vacuum $|0\ra_\out$, using an $S$-matrix $U=e^{-iH}$~\cite{Israel1976}  
\be |0\ra_{\rm out} =
e^{-iH}|0\ra_{\rm in} \label{map}\;,
 \ee
with
\be H = \sum_{k=-\infty}^{\infty}  H_k=\sum_{k=-\infty}^\infty ig_k(a_k^\dagger b_{-k}^\dagger - a_k b_{-k})  \;. \label{ham}\ee
The coefficients $\alpha_k$ and $\beta_k$ are obtained in terms of $g_k$ using the Baker-Campbell-Hausdorff theorem as 
\be
\alpha_k^2 &=& \cosh^2 g_k\;, \label{alpha2}\\
\beta_k^2 & = & \sinh^2 g_k \;. \label{beta2}
\ee
with $\alpha_k^2-\beta_k^2=1$. Standard arguments~\cite{Hawking1975} enforcing analyticity on the solution to the free-field equations 
give
\be
\alpha_k^2 = e^{\omega/T}\beta_k^2\;, \label{hrelation}
\ee
as usual, where $\omega = |k|$ is the frequency of mode $k$ and $T$ is the black hole temperature.

Let us study the vacuum state in more detail, in preparation for the calculations to follow.  Using (\ref{ham}), we can write
\be
|0\ra_\out =\prod_{k=-\infty}^{\infty} e^{g_k (a_k^\dagger b_{-k}^\dagger -a_k b_{-k})}|0\ra_\inv\;.
\ee
Each factor (for each mode), can be rewritten using the disentangling theorem for SU(1,1) (see, e.g., the Appendix in \cite{Wodkiewicz1985})
\be
e^{g_k(a_k^\dagger b_{-k}^\dagger - a_k b_{-k})} = e^{G a_k^\dagger b_{-k}^\dagger} e^{F (a_k^\dagger a_k + b_{-k}b_{-k}^\dagger )} e^{-G a_k b_{-k}}\;,\ \ \ \ \ 
\ee
where
\be
G &=& \tanh g_k= \frac{\beta_k}{\sqrt{1+\beta_k^2}}\;,\\
F & =& -\ln \cosh g_k = \ln {\frac1{\alpha_k}}\;,
\ee
so that (for $|0\ra_\inv=|0\ra_a|0\ra_b$)
\be
|0\ra_\out
=\prod_k \frac1{\cosh^2 g_k}\sum_{n_k,n_k^\prime} e^{-(n_k+n_k^\prime)\omega/2T}
|n_k,n^\prime_{-k}\ra_a|n^\prime_{k},n_{-k}\ra_b \nonumber\\ \! \! \!
\ee
where $|n_k\ra$ is an $n$-particle state and $|n_{-k}\ra$ a state with $n$ antiparticles. We can then calculate the density matrix of the outgoing radiation by tracing over sector II:
\be
\rho_{\rm I} = {\rm Tr}_{\rm II} |0\ra_\out \la 0| = \prod_k \rho_k \otimes \rho_{-k}\label{hawking}
\ee
where  
\be
\rho_k& =&  \frac1{1+\beta_k^2} \sum_{n=0}^\infty \left(\frac{\beta_k^2}{1+\beta_k^2}\right)^n |n_k\ra \la n_k| \nonumber \\
&=&(1-e^{-\omega/T}) \sum_{n_k=0}^\infty e^{-n_k \omega/T} |n_k\ra \la n_k|\nonumber \label{mat} \\
\ee
and similarly for $\rho_{-k}$, the density matrix for antiparticles. These are the standard density matrices for spontaneous emission of radiation from a black hole~\cite{Bekenstein1975,BekensteinMeisels1977}. 
The mean number of particles emitted in region I (mode $a$) is also easily
calculated from (\ref{bog1}) and (\ref{map}) as
 \be
  \la N_{\rm I}\ra =\sum_{k=-\infty}^{\infty}
{_ \out}\la 0| a_k^\dagger a_k|0\ra_\out 
=\sum_{k=-\infty}^{\infty}\beta_k^2\;, 
\ee
which is the famous Planck distribution of Hawking radiation:
\be
\beta_k^2=\frac{e^{-\omega/T}}{1-e^{-\omega/T}}\;.
\ee
From (\ref{mat}) we can calculate the von
Neumann entropy of  frequency mode $k$ of the inside or outside region as
\be 
S(\rho_k) = -\Tr \rho_k\log \rho_k= \frac{\omega/T}{e^{\omega/T} - 1} +
\log(1-e^{-\omega/T})\label{ent}\;. \ \ \ \ \ 
\ee 
This entropy is perfectly thermal, which implies that it does not depend on the distribution
of matter that formed the black hole.

\section{Stimulated Emission}
If matter is incident on the black hole during its formation (as is inevitably the case), what is its fate? Here we show (in accordance with previous work~\cite{Wald1976,BekensteinMeisels1977,PanangadenWald1977,Jones1985,Sorkin1987,AudretschMueller1992,Schiffer1993,MuellerLousto1994}), that the black hole stimulates emission of radiation in the same mode as it entered (except for a redshift that we do not consider here, but discuss further below). 

Let us consider a single mode $k$, with $m$ particles incident in region I, just outside the black hole horizon (we treat the case of particles traveling just inside the horizon, that is, $b$ modes, later). Then (we now use $|\psi\ra_\out$ to denote the outgoing state)
\be
|\psi\ra_\out=e^{-iH_k}|m\ra_a|0\ra_b\;,
\ee
and the density matrix in sector I becomes, using (\ref{ham}) and the disentangling theorem just as before
\be
\rho_{\rm I}=\Tr_{\rm II} |\psi\ra_\out \la \psi |=\rho_{k|m}\otimes \rho_{-k|0}\;.
\ee
Here, $\rho_{-k|0}$ is the density matrix of antiparticles given that no antiparticle was incident in mode $k$, and $\rho_{k|m}$ is the density matrix of mode $k$ given that $m$ particles were incident in that mode:
\be
\rho_{k|m}&=&\frac1{(1+\beta_k^2)^{m+1}}\times \nonumber\\
&& \sum_{n=0}^\infty\left(\frac{\beta_k^2}{1+\beta_k^2}\right)^{n}{{m+n}\choose{n}}|m+n\ra\la m+n|\;,\nonumber \\
&& 
\ee
which can be rewritten as 
\be
\rho_{k|m}=\sum_{n=0}^\infty p(n|m)|n\ra\la n| \label{denmat}
\ee
with the conditional probability
\be
p(n|m)=\frac1{(1+\beta_k^2)^{m+1}}\left(\frac{\beta_k^2}{1+\beta_k^2}\right)^{n}{{m+n}\choose{n}}\;.\ \ \ \ \ 
\ee
As the density matrix of modes $k'\neq k$ for which there is no incident particle factorizes, we from now on focus only on the incident mode and omit the label $k$ for clarity. Writing $z=\beta^2/(1+\beta^2)=e^{-\omega/T}$, the conditional probability to observe $n$ particles (both spontaneous and stimulated) in region I given that $m$ particles where absorbed in region II during the formation of the black hole is 
\be
p(n|m)=(1-z)^{m+1}z^{n}{{m+n}\choose{m}}\;. \label{condprob}
\ee
As a consequence, the mean number of particles emitted into region I is
\be
 \la N_{\rm I}\ra &=&
{_\out}\la \psi| a_k^\dagger a_k|\psi\ra_\out = {_{a,b}}\la m,0 |\alpha^2 a_k^\dagger a_k +\beta^2 b_{-k} b_{-k}^\dagger|m,0\ra_{a,b} \nonumber \\
&=&(1+\beta^2)m +\beta^2\;, \label{number}
\ee
that is, the $m$ incident particles have stimulated the emission of $(1+\beta^2)m$ such particles outside of the horizon in region I (in agreement with previous work~\cite{Sorkin1987,AudretschMueller1992,MuellerLousto1994}). Note that as $\beta^2 m$ anti-particles are stimulated behind the horizon in region II, particle number is conserved. We should also point out that because the incident particle carries energy and momentum, the black hole does not have to donate mass in order to allow the emission of stimulated pairs, as it does for virtual pairs. 

Do the particles stimulated in region I carry the information that was inherent in the particles that were absorbed during the formation of the black hole? To study this question, we have to turn to quantum information theory~\cite{Peres1993,Preskill1998,NielsenChuang2000,Wilde2013}, the theory that deals with the transmission of information through quantum channels. Here, we are interested in the classical information encoded in the $m$ incident particles. The capacity of quantum channels to transmit classical information is known as the ``Holevo capacity" in the literature~\cite{SchumacherWestmoreland1997}, which we briefly review here. 

Let us imagine that the particles incident on the horizon in the formation of the black hole (so-called ``early-time" modes) 
had been ``prepared", that is, a copy of those states had been retained. As we are considering the fate of classical information here, this is always possible. We can imagine that information is encoded in such states via a so-called ``dual-rail" encoding, where a logical ``0" is encoded by sending a single particle, and a logical ``1" is encoded by sending a single anti-particle (there are other ways to encode information in particle states, but our conclusions do not dependent on the encoding, only the rate of information transfer will). The question now is whether an observer at late-time, measuring particles and anti-particles in region I, can determine a message encoded by the preparer. This question is answered in the affirmative if the Holevo capacity of the channel is non-vanishing. 
Note that it is not necessary for information preservation that a detector can unambiguously identify a logical zero from a logical one. Indeed, in noisy channels signals can easily be confused. But the classical theory of information has taught us that information can be perfectly preserved using standard methods of error correction, as long as the capacity of the channel is positive~\cite{CoverThomas1991}. Of course, error correction is necessary to protect messages from noise, which entails encoding the information into codewords that are typically longer then the encoded message. If a message of $n$ bits is encoded into $n+m$ bits that are sent through the channel and $n$ bits are perfectly retrieved, have we not lost $m$ bits of information anyway? The answer is no: these $m$ ``lost bits" are not information, they are entropy. Losing entropy does not violate any laws of physics because entropy cannot be used for signaling. Thus, information transmission is really an ``all-or-nothing" case: if the capacity vanishes, then no amount of error correction could possibly save the message. If the capacity is non-zero, then {\em all} of the information can be recovered in principle (but it may be less than the maximum in practice, unless error correcting codes appropriate to the noise level are used). 

Classical information is preserved in the formation of the black hole if the message encoded by the preparer (and sent into the forming black hole at early time) can be decoded by an observer in region I outside of the black hole horizon, at late time. The information shared between the preparer $X$ and the radiation field outside the black hole is the mutual entropy between them:
 \be
H(X:{\rm I}) = S(\rho_{\,\rm I}) +S(\rho_X)-S(\rho_{\,{\rm I},X})\;, \label{mutent}
\ee
where $S(\rho_I)$ is the von Neumann entropy in region I, $S(\rho_X)$ is the von Neumann entropy of the preparer, and $S(\rho_{\,{\rm I},X})$ is the joint von Neumann entropy between the preparer and the radiation field. We remark that if we found the radiation field outside of the horizon at late time {\em independent} from the preparer, the shared entropy (\ref{mutent}) would vanish, and the preparer's information would be irretrievably lost. The capacity of the channel $\chi$ is the maximum of the shared entropy over the probability distribution of the signal states~\cite{SchumacherWestmoreland1997}:
\be
\chi=\max_p H(X:{\rm I})\;.
\ee

The preparer's entropy is easily calculated. Let us assume she uses state `0' with probability $p$ and state `1' with probability $1-p$. Then 
\be
S(\rho_X)=-p\log p-(1-p)\log(1-p)\equiv H[p]\;.
\ee
The preparer's entropy is nothing but the entropy of the signal, and this is maximized at one bit for $p=1/2$. 

The density matrix in region I, $\rho_{\rm I}$, is a probabilistic mixture of the density matrix that arises if a logical `0' or a `1' was sent. If $X$ intends to send a `0', she sends a single particle into the forming black hole, giving rise to the matrix in region I
\be
\rho_{\rm I}(0)=\rho_{k|1}\otimes\rho_{-k|0}\;,
\ee
while if she encodes a `1', she sends an anti-particle, which leads to
\be
\rho_{\rm I}(1)=\rho_{k|0}\otimes\rho_{-k|1}\;.
\ee
Thus, in response to the preparer's actions the density matrix in region I at late times is
\be
\rho_{\rm I}(p)=p\rho_{\rm I}(1)+(1-p)\rho_{\rm I}(0)\;. \label{mixed}
\ee
The joint density matrix $\rho_{\,{\rm I},X}$ of preparer and region I is block-diagonal
\be
\rho_{\,{\rm I},X}=\sum_{i=0}^1 p(i) \rho_{\rm I}(i) |i\ra\la i|\;, \label{joint}
\ee
where $|i\ra$ are the preparer's basis states. 

\begin{figure}[htbp] 
   \centering
   \includegraphics[width=4.4in]{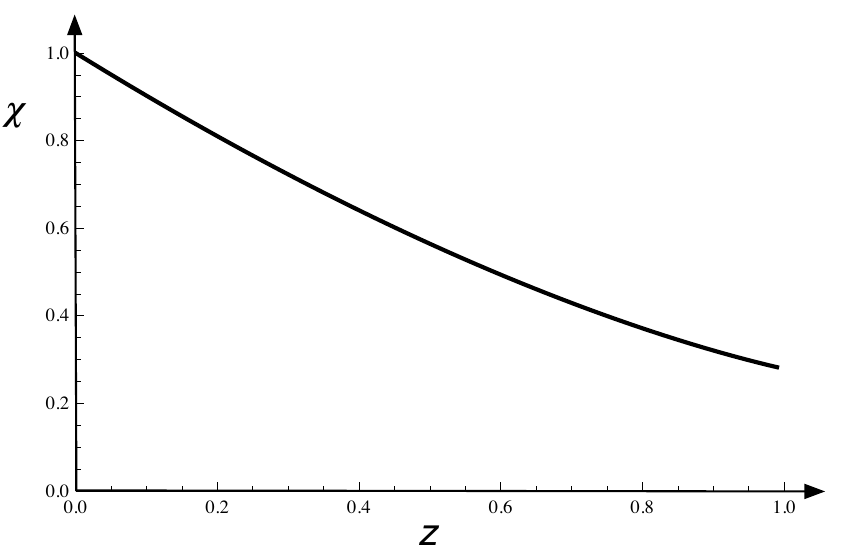} 
   \caption{The Holevo capacity of the information transmission channel for early-time particles that propagate just outside the horizon of the forming black hole, as a function of $z=e^{-\omega/T}$. The capacity to transmit information behind the horizon, using the ``mirror" mode $b$ propagating just inside the horizon, is the same. }
   \label{holevocap}
\end{figure}

We can now calculate the entropies in Eq.~(\ref{mutent}). Owing to the block-diagonal nature of (\ref{joint}), the entropy of the joint system (preparer and region I) is easily calculated as
\be
S(\rho_{\,{\rm I},X})=H[p]+\sum_i p(i) S(\rho_{\rm I}(i)) \;.
\ee
Now, because the particle and anti-particle density matrices factorize, we have
\be
S(\rho_{\rm I}(0))&=&S(\rho_{k|1})+S(\rho_{-k|0})\;,  \nonumber \\
S(\rho_{\rm I}(1))&=&S(\rho_{k|0})+S(\rho_{-k|1}) \;.
\ee
Particle and anti-particle entropies are identical here: $S(\rho_{k|1})=S(\rho_{-k|1})$ and $S(\rho_{k|0})=S(\rho_{-k|0})$. Using (\ref{denmat}) and (\ref{condprob}), we find
\be
S(\rho_{k|0})&=&-\log(1-z)-\frac z{1-z}\log z\;,\\
S(\rho_{k|1})&=&-2\log(1-z)-\frac{2z}{1-z}\log z-(1-z)^2\Delta\;,\ \ \ \ \ 
\ee
where
\be
\Delta=\sum_{m=0}^\infty z^m(m+1)\log(m+1)\;.
\ee
As a consequence
\be
S(\rho_{\,{\rm I},X})=H[p]-3\log(1-z)
&-&\frac{3z}{1-z}\log z-(1-z)^2\Delta\;. \nonumber \\
& &
\ee
To calculate (\ref{mutent}), we still need to calculate the entropy of the mixed density matrix (\ref{mixed}), which depends on $p$. 
Because the density matrix 
\be
\rho_{\rm I}(p)&=&(1-z)^3 \times \\
&&\sum_{m,m'=0}^\infty z^{m+m'-1}\Bigl( p\,m'+(1-p)m\Bigr) |mm'\ra\la mm'| \nonumber
\ee
is symmetric under the replacement $p\to1-p$, the mutual entropy (\ref{mutent}) is maximized at $p=1/2$, and it is sufficient to calculate this case. We find
\be
S(\rho_{\rm I}(1/2))&=&1-3\left(\log(1-z)+\frac{z}{1-z}\log z\right) \nonumber \\
&-&\frac12(1-z)^3\sum_{m=0}^{\infty}z^m(m+1)(m+2)\log(m+1)\;.
\ee
Collecting terms, we obtain the capacity 
\be
\chi&=&1-\frac12(1-z)^3\sum_{m=0}^{\infty}z^m(m+1)(m+2)\log(m+1) \nonumber \\
&+& (1-z)^2 \sum_{m=0}^\infty z^m(m+1)\log(m+1)\;,\label{holcap}
\ee
shown in Fig.~\ref{holevocap}. Note that this capacity is formally identical to the recently calculated Holevo capacity of the Unruh channel~\cite{Bradler2011}, which describes the rate of information transmission between a stationary Minkowski signaler and a uniformly accelerated observer, except that in that case the surface gravity $\kappa$ is replaced by the proper acceleration $r$ in the definition of the temperature. This identity is of course not surprising, given the close relationship between Unruh radiation and Hawking radiation~\cite{BirrellDavies1982}. 

Fig.~\ref{holevocap} indicates that at $z=0$ ($T\to0$), information can be retrieved at the rate that it is produced. For example, if information is sent into the channel at 1bit/s, information can be retrieved at that rate. At $z=e^{-\omega/T}=0.5$ , the capacity has dropped to about 0.56, which indicates that the rate at which information can be retrieved with perfect accuracy has dropped to 0.56 bits/s (not, as is sometimes remarked, that only 56\% of the information can be retrieved). In other words, at $z=0.5$ it will take at least 1/0.56 seconds to retrieve one bit. This slowdown is due to the necessary error-correcting encoding of the signal stream that renders communication in noisy channels possible. As is clear from Fig.~\ref{holevocap}, perfect information transmission is possible even in the limit $z=1$, that is, $T\to \infty$. Note that had we used Hawking's result Eq.~(\ref{hawking}) to calculate the capacity, $\chi=0$ exactly.

In the present calculation, we have ignored any red-shift factors that must accompany the evolution of particles towards future null infinity.  This can often be done within an information-theoretic treatment because each frequency mode $k$ is shifted in a unique way (if the interactions do not mix modes, as in this 1+1-dimensional model). From a practical point of view, the observer at future null infinity simply should adjust his detector to account for the frequency shift. The treatment of stimulated emission including the red shift is straightforward, but more cumbersome~\cite{Wald1976,PanangadenWald1977,AudretschMueller1992,MuellerLousto1994}. Note that the redshift reduces the capacity of the channel further because the (exponentially) decreased frequency of the signal slows down the rate at which particles can be recorded. However, as long as the capacity is finite and the signal not infinitely redshifted, perfect information transmission is still possible.

We could have equally well calculated the capacity of information transmission from the outside to the inside of the black hole, using the particles traveling just {\em inside} the event horizon of the forming black hole (modes $b$ in Fig.~\ref{default}b). Indeed, at past null infinity the $b$ modes and the $a$ modes overlap. Particles traveling just inside the event horizon stimulate the emission of anti-particles in region I~\cite{AudretschMueller1992,MuellerLousto1994,AdamiverSteeg2006}, and because of the dual-rail encoding the capacity is the same: we can decode the signal simply by flipping all the zeros into ones. Thus, observers both inside and outside the black hole horizon can in principle decode Alice's message with perfect accuracy.  

Still, the present treatment is unsatisfactory because it does not deal with the absorption and reflection of particles (that is, with grey-body factors), which are important when studying signaling in the presence of an already-formed black hole. Indeed, it is clear that in the formalism we just described, matter does not interact with (is scattered by) the black hole: particles that travel just inside the event horizon (the $b$-modes) remain there, while particles just outside the horizon (the $a$-modes) remain outside. Indeed, within the standard semi-classical approach an $S$-matrix for matter interacting with a black hole has not yet been found (but candidate interactions exist in black hole descriptions based on string theory, see, e.g.,~\cite{Witten1992}). 

We can extend the treatment of this section to create an effective $S$-matrix that correctly describes the absorption, reflection, and stimulated emission of {\em late-time} modes, that is, particles that interact with the black hole after its formation, using a trick due to Sorkin~\cite{Sorkin1987}. In this formalism, we can directly study the fate of information that is incident on a static black hole.

\section{Stimulated emission from late-time modes}
As we have just seen, stimulated emission due to particles present just before formation of the black hole (early-time modes) implies that the information that gave rise to the black hole is not lost in principle, but within this formalism we cannot follow the fate of information incident on the black hole after its formation, by  {\em late-time} modes. To describe such modes, we follow Sorkin~\cite{Sorkin1987}. Late-time modes (which we denote by $c$)\footnote{Note that we have changed the nomenclature of Sorkin, in order to keep with our previous definition of $a$ and $b$ modes.} are exponentially blue-shifted with respect to the early-time modes $a$ and $b$ that are the positive and negative-frequency parts of matter and radiation that propagate just outside and just inside the horizon introduced earlier. Because of this strong blue-shift, $a$ and $b$ modes are unexcited in the quantum field corresponding to the late-time modes, and therefore the support of the fields $\phi_a,\phi_b$, and $\phi_c$ is for all practical purposes disjoint~\cite{Sorkin1987}. As a consequence, the creation and annihilation operators associated with mode $c$ commute with those of $a$ and $b$. These modes are indicated schematically in Fig.~1b.

The Bogoliubov transformation connecting the outgoing mode $A_k$ to $a_k$ and the late-time mode $c_k$ can be taken as~\cite{Sorkin1987}: 
\be A_k &=& e^{-iH}a_k e^{iH} = \alpha
a_k - \beta b^\dagger_{-k}+ \gamma c_k\;,\label{bogol} \ee
with $\alpha^2-\beta^2+\gamma^2=1$ due to unitarity.

It is possible to amend the Hermitean operator $H$ that implemented the Bogoliubov transformation (\ref{bog1}-\ref{bog2}) to give rise to (\ref{bogol}):
 \be
 H = \sum_{k=-\infty}^\infty ig_k(a_k^\dagger b_{-k}^\dagger - a_k b_{-k} )+ig^\prime_k (a_k^\dagger c_k- a_k c_k^\dagger)  \;.\ \ \ \  \ \ \label{ham2}
 \ee
While the standard term Eq.~(\ref{ham}) describes the entanglement between modes across the event horizon, 
the extra term describes scattering of late-time modes $c_k$ by the static black hole horizon, with an interaction strength $g_k^\prime$.
Such an interaction is entirely natural within quantum field theory and is completely analogous to Hamiltonians used in
quantum optics to describe transformations in linear optics~\cite{Leonhardt2003}. In fact, the first term (\ref{ham2}) corresponds to the Hamiltonian of an {\em active} optical element leading to state squeezing with strength $g_k$, while the second one describes
a {\em passive} element: a simple beam splitter with interaction strength $g_k^\prime$. We will discuss the magnitude of this coupling constant in relation to $g_k$ below. 
 
Using (\ref{ham2}) in (\ref{bogol}), the Bogoliubov coefficients
$\alpha$, $\beta$, and $\gamma$ can be expressed in terms of $g_k$
and $g^\prime_k$ as 
\be
\alpha ^2 &=& \cos^2(\gp_k w) \label{alpha}\;,\\
\beta^2&=&(\frac{g_k}{\gp_k})^2\frac{\sin^2(\gp_k w)}{w^2}\; \label{beta}\;,\\
\gamma^2&=&\frac{\sin^2({\gp_k w})}{w^2} \label{gamma}\;,
\ee 
where
$w^2=1-(g_k/\gp_k)^2$. 

We can now calculate the reduced (marginal) density matrix of outgoing
radiation in region I when no particles are incident (notation
$k|0$) just as before, that is, construct (we focus on a single mode $k$ again, and denote the outgoing state with $|0\ra_\out$ )
\be
|0\ra_\out=e^{-iH_k}|0\ra_a|0\ra_b|0\ra_c\;,
\ee
and calculate the density matrix 
\be
\rho_{\rm I}=\Tr_{\rm II}|0\ra_\out\la 0| 
\ee
by tracing the full density matrix over region II (containing both modes $b$ and $c$). As in the previous cases, the anti-particle density matrix $\rho_{-k|0}$ factorizes, and we find 
\be 
\rho_{k|0}=\frac1{1+\beta^2}\sum_{n=0}^\infty
\left(\frac{\beta^2}{1+\beta^2}\right)^{n} |n\ra\la n|\;. \label{rho0}
 \ee
This expression when written in terms of $\beta^2$ is formally identical to expression 
(\ref{mat}), except that now 
\be
\beta^2= \frac{\Gamma}{e^{\omega/T}-1}\;,
\ee
where $\Gamma$ is the absorptivity of the black hole, with $\Gamma=1-\gamma^2$.
Thus, (\ref{rho0}) is just the standard density matrix of spontaneous emission including grey-body factors.  It reduces to (\ref{mat}) in the limit $\Gamma\rightarrow1$. 

Using the aforementioned expression for $\beta^2$ along with the relation  $\beta^2=e^{-\omega/T}\alpha^2$ (which is due to the standard analyticity argument for the solution to the field equations that carries through as before) we also identify
\be
\alpha^2=\frac{\Gamma}{1-e^{-\omega/T}}\;,
\ee
owing to $\alpha^2-\beta^2+\gamma^2=1$.

Consider now the outgoing state $|\psi\ra_\out$ when $m$ late-time particles are incoming (all in the same mode $k$), constructed as before via
 \be
|\psi\ra_\out = e^{-iH_k}|m\ra_\inv\;, \label{state}
 \ee 
with $H_k$ from (\ref{ham2}) and where $|m\ra_\inv=|0\ra_a|0\ra_b|m \ra_c$, i.e., the state with $m$ incoming particles in mode $c$ on ${\mathscr I}^-$.
From Eq.~(\ref{bogol}), we immediately obtain the number of
particles emitted into mode $a$ if $m$ were incident in mode $c$, since
 \be {_\out}\la \psi|a_k^\dagger a_k|\psi\ra_\out = \gamma^2 m +\beta^2\;. \ \ \ \ \ \ \label{partnum}
\ee 
This shows that, in addition to the standard spontaneous emission term $\beta^2$, 
detectors at future null infinity record $\gamma^2 m$ particles. The latter comprise two contributions: $(1-\alpha^2) m$ due to elastic scattering with a quantum absorption probability $\alpha^2$,  and $\beta^2 m$ particles due to stimulated emission, as before in Eq.~(\ref{number}). Just as in the early-time case, particle number is conserved overall, and stimulated emission does not cost the black hole any additional energy. 

The density matrix of outgoing radiation in region I when $m$
particles are incident (notation $k|m$) can be calculated from
$|\psi\ra_\out\la \psi|$ by repeated application of the disentangling
theorems for SU(2) and SU(1,1), and tracing over the degrees of
freedom of region II (modes $b$ and $c$). When no antiparticles
are sent in, the antiparticle part of the density matrix factorizes again and we can
write $\rho_{\rm I}=\rho_{k|m}\otimes\rho_{-k|0}$ with  
\be
\rho_{k|m}=\sum_{n=0}^{\infty}p(n|m)|n\ra\la n|\;,
\label{mpartmat} 
\ee 
with
\be
p(n|m)&=&Z_m^2\sum_{l=0}^\infty\frac{n!\,l!}{m!(l-m+n)!} 
 \left[\sum_{k=0}^{\min(n,m)}(-1)^k {m \choose k}{{l-m+n}\choose{n-k}}\Bigl(\frac{\gamma^2}{\alpha(1+\alpha)}\Bigr)^k\right]^2\nonumber \\
&\times&  \Bigl[\frac{\beta^2(1+\alpha)^2}{(1+\alpha+\beta^2)^2}\Bigr]^n
  \Bigl[\frac{\beta^2\gamma^2}{(1+\alpha+\beta^2)^2}\Bigr]^l  \label{myexp}\nonumber \\
\ee
and
\be
Z_m^2=\left(\frac{1+\alpha}{1+\alpha+\beta^2}\right)^{2}\left(\frac{\alpha^2(1+\alpha)^2}{\gamma^2}\right)^m\;.
\ee

This expression for $p(n|m)$, the probability
to detect $n$ outgoing particles at ${\mathscr I}^+$ if $m$ were
incident on the static black hole, can be rewritten using a resummation technique described by Panangaden and Wald~\cite{PanangadenWald1977} to read
\be
p(n|m)=R_{nm}\sum_{k=0}^{\min(n,m)}(-1)^k{{m}\choose {k}}{{m+n-k}\choose {n-k}}\left(1-\frac{\gamma^2}{\alpha^2\beta^2}\right)^k \label{BM}
\ee
with 
\be
R_{nm}=\frac1{1+\beta^2}\left(\frac{\beta^2}{1+\beta^2}\right)^{m+n}\left(\frac{\alpha^2}{\beta^2}\right)^m\;,
\ee
which agrees exactly with the conditional probability obtained by Bekenstein and Meisels~\cite{BekensteinMeisels1977} using maximum entropy methods, and by Panangaden and Wald~\cite{PanangadenWald1977} in quantum field theory (but using a very different method than ours).

We can use our expression for $p(n|m)$ to put limits on the scattering strength $\gp_k$.
For example, the probability that no particle is found outside the horizon while 
a single quantum is incident is given by
\be p(0|1)=\frac{\alpha^2}{(1+\beta^2)^2}=\frac{\alpha^2}{(1+\alpha^2e^{-\omega/T})^2}\;,
\ee
which tends to the single-quantum absorption probability $\alpha^2$ in the zero-temperature limit. Thus, while the parameter $g_k$ relates the mode frequency to the temperature of the black hole, $\gp_k$ in turn sets the reflectivity of the grey body via 
\be
(\gp_k/g_k)^2=1+\frac{1-\alpha^2}{\alpha^2}e^{\omega/T}
\ee
Because $0\leq p(0|1)\leq1$ for all temperatures, we can deduce that
$0\leq\alpha^2\leq1$, which in turn implies $\gp_k\geq g_k$ or else $\gp_k=0$ [see Eq.~(\ref{alpha})]. In other words, if there is to be scattering from the black hole horizon, the coupling $g_k$ sets a lower bound for the magnitude of the scattering effect.
The solution $\gp_k=0$ corresponds to the familiar Hawking transformation (\ref{ham}) where $\Gamma=\alpha^2=1$, but in hindsight cannot account for a thermodynamically consistent picture that includes scattering from the potential barrier surrounding the black hole, in which  $\Gamma<1$, as pointed out by Bekenstein and Meisels~\cite{BekensteinMeisels1977} (and reiterated by Bekenstein~\cite{Bekenstein1993}). 

An interesting choice for the scattering rate is $\gp_k=g_k$, that is, the magnitude of scattering is set by the parameter $g_k$ that describes the strength of entanglement between modes across the horizon. This limit describes perfect absorption ($\alpha^2=1$) where we find
\be
p(n|m)={{m+n}\choose{m}}\frac{e^{-n\omega/T}}{(1+e^{-\omega/T})^{n+m+1}}\;.
\ee
But even in this extreme case, the classical (effective) absorption probability
$\Gamma=\alpha^2(1-e^{-\omega/T})$ is strictly smaller than unity due to the stimulated emission effect, thus allowing for the preservation of classical information.

Let us now calculate the capacity for the black hole channel when information is sent in after the formation of the black hole, that is, via late-time modes. We illustrate this again in the channel with ``dual-rail" encoding, that is, a logical `0' is encoded by sending in a particle, while a logical `1' is encoded by sending an anti-particle.  The probability to observe $n$ particles when one particle is sent in at late-time in mode $c$ is [from Eq.~(\ref{myexp}) or (\ref{BM})]
\be
p(n|1)=\frac{\alpha^2}{(1+\beta^2)^2}\left(\frac{\beta^2}{1+\beta^2}\right)^n(1+n)\;,
\ee
which is again formally identical to the expression from early-time stimulated emission, but there is a crucial difference. Because $\beta^2=\alpha^2e^{-\omega/T}$, the limit of perfect absorption ($\alpha^2\to1$) in terms of $z=e^{-\omega/T}$ is
\be
p(n|1)\to  \frac1{(1+z)^2}\left(\frac{z}{1+z}\right)^n(1+n)\;,
\ee
compared to 
\be
p(n|1)= (1-z)^2z^n(1+n)
\ee
in the case of early-time stimulated emission, as can be seen from Eq.~(\ref{condprob}). Thus, late-time stimulated emission (in the perfect absorption limit)  is obtained from the early-time stimulated emission probability simply by the transformation $z\to z/(1+z)$. This result holds generally for arbitrary $m$, that is, an arbitrary number of incoming particles. As a consequence, we do not have to recalculate the capacity of the black hole channel, as it can be obtained from Eq.~(\ref{holcap}) by replacing $z$ with $\zeta=z/(1+z)$. For the numerical values of the capacity, we see that the capacity is larger than for the early-time modes, as it reaches $\chi\approx 0.6$ at $z=1$ (the value for $z=0.5$ in the early-time channel, see Fig.~\ref{holevocap}). 

\section{Discussion}
When discussing the fate of classical information in black hole formation and evaporation, authors usually take on one of several positions (see, for example,~\cite{Preskill1992,Giddings1996,Mathur2012}): some think information is indeed irretrievably lost, and black holes point towards new physics that is fundamentally different from the physics we know. Others believe that information will ultimately leave the black hole during its evaporation, either gradually or mostly during the last stages of a black hole's life. Others, again, think that information could be permanently stored either in a Planck-sized remnant, or even a baby universe. Very few protagonists seem to have considered that information is never lost within a black hole to begin with because a classical copy of the information remains outside the event horizon. At first sight, such a position seems preposterous: after all, how could one doubt that matter that interacts with a black hole will be absorbed by it, and all of the classical and quantum numbers along with it? But this point of view fails to make a distinction between information and the matter (or radiation) that carries it. Information is not a quantity that is irrevocably linked to a particle, such as spin or charge. Information can be transferred from one particle to another, for example, it can be copied (although quantum information can only be copied imperfectly~\cite{Dieks1982,WoottersZurek1982}). Stimulated emission of radiation is precisely the process of copying information: one particle comes in, two leave with the same exact set of quantum numbers. Now, because of the aforementioned no-cloning theorem, quantum information cannot be copied perfectly, but it turns out that the spontaneous emission of radiation (the Hawking radiation of the black hole) conveniently prevents perfect cloning by supplying the necessary minimum amount of noise~\cite{AdamiverSteeg2006}. Classical information, fortunately, can still be copied and perfectly retrieved even in the presence of Hawking noise, via standard methods of channel coding.

It is perhaps surprising that the problem of information transmission within the context of black holes has remained mysterious for so long, in particular given the fact that the stimulated emission process for black holes has been discussed numerous times~\cite{Wald1976,BekensteinMeisels1977,PanangadenWald1977,Jones1985,Sorkin1987,AudretschMueller1992,Schiffer1993,MuellerLousto1994}. We believe that the reason that stimulated emission was not viewed immediately as the solution to the information paradox is due to the fact the quantum information theory, and the concomitant understanding of the fate of classical and quantum information over quantum channels, is a relatively new addition to the canon of physics, and in particular is not part of the standard vocabulary of workers in the field of quantum gravity. But even the classical theory of information was not generally used to understand the fate of information interacting with black holes.  For example, Schiffer discussed stimulated emission in the presence of black holes in 1993~\cite{Schiffer1993}, but concluded that ``Unfortunately this mechanism is not efficient enough to resolve the black-hole information paradox because the thermal radiation overpowers stimulated emission of bosons for the majority of modes." The thermal radiation Schiffer refers to here is of course Hawking radiation, which represents the noise inherent in the channel.  But noise cannot overpower an information signal unless the capacity of the information transmission channel is zero. If the capacity is non-vanishing, classical information can be perfectly retrieved as long as it is encoded within an appropriate error-correcting code. What the calculations we present here show is that the capacity of the quantum black hole channel to transmit classical information is decidedly non-vanishing. And even though the phenomenon of red shift of early-time particles was not considered when early-time accretion was considered, this red shift is only expected to lower the information transmission capacity, not to abolish it. 

We have not here addressed the fate of quantum information that crosses the black hole horizon, or the capacity to reconstruct the {\em quantum} states that formed the black hole. The former (signaling with quantum states) is possible in principle, as has been discussed for quantum communication in Rindler space~\cite{Bradleretal2012}. A calculation of the quantum capacity of a black hole should also shed more light on current controversies that are centered on understanding the properties of entanglement between signalers, observers and the black hole, black hole complementarity~\cite{Susskindetal1993,Stephensetal1994}, and whether quantum ``firewalls"~\cite{AMPS2012} are necessary to prevent a violation of the no-cloning theorem. Finally, we may learn about the ultimate fate of the equivalence principle in the light of quantum information. 


We thus conclude that black holes are, information-theoretically speaking, fairly ordinary black bodies that scramble the signals we may endeavor to send, but conserve information after all.  In hindsight it is perhaps ironic that Hawking radiation, which for many years was viewed as the herald of a fundamental violation of the laws of quantum mechanics, is instead instrumental in preserving the integrity of quantum mechanics, because it upholds the no-cloning theorem.

\ack
We are grateful to N.J. Cerf and C.O. Wilke for crucial discussions and comments on
the manuscript. We also thank K. Br\'adler, P.C.W. Davies, C.G. Callan, J.P. Dowling, D. Gottesman, G.M. Hockney, P. Kok, H.
Lee, and U. Yurtsever for discussions. This work was carried out
in part at the Jet Propulsion Laboratory (California Institute of
Technology) under a contract with NASA, with support from the Army
Research Office's grant \# DAAD19-03-1-0207.

\section*{References}

\end{document}